\documentclass[a4paper]{jpconf}
\usepackage{graphicx}
\usepackage{amssymb}
\usepackage{amsmath}

\newcommand{\SRO}{Sr$_2$RuO$_4\:$}

\def\bk{{\bold{k}}}
\def\up{{\uparrow}}
\def\down{{\downarrow}}
\def\dg{{^\dagger}}

\begin{document}
\title{A note on the upper critical field of \SRO under strain}

\author{Aline Ramires$^1$ and Manfred Sigrist$^2$}

\address{$^1$ Institute for Theoretical Studies, ETH Z\"{u}rich, CH-8092, Z\"{u}rich, Switzerland}

\address{$^2$ Institute for Theoretical Physics, ETH Z\"{u}rich, CH-8093, Z\"{u}rich, Switzerland}

\ead{aline.ramires@eth-its.ethz.ch}

\begin{abstract}
In the light of the recently discussed mechanism for suppression of superconductivity in multi-orbital systems called \emph{inter-orbital effect}, here we extend our analysis of the upper critical field in \SRO now under strain. We show that the presence of the standard orbital effect and the new mechanism can consistently account for the qualitative changes observed in the upper critical field for the strained system, in particular, combining the overall enhancement of the critical field with the reduction in the anisotropy. The proposed picture holds for a triplet superconducting state, showing that a singlet state is not the only possibility to account for the observations. We suggest further experiments in order to clarify our understanding about the superconducting phase of \SRO\!\!.
\end{abstract}

\section{Introduction}\label{SecInt}

\SRO has been extensively explored experimentally since it is one of the few putative chiral triplet superconductors \cite{Mac}. The actual character of the order parameter and the origin of the pairing mechanism in this material are still under debate. The lack of agreement in the community steams mainly from the fact that this is a multi-band system, and different bands may favor distinct pairing mechanisms and consequently order parameters. One mechanism suggests the dominance of the $\gamma$-band due to its proximity to a van-Hove singularity \cite{Nom1, Nom2, Yan, Nom3, Wan}, while another focuses on the enhancement of spin and charge fluctuations due to nesting \cite{Rag, Chu, Fir} in the almost 1D $\alpha$- and $\beta$-bands. The dominance of one over the other is not very clear on a theoretical level and might be dependent on the calculation scheme chosen. Recent studies, based on the renormalization group technique, considering all three bands on the same footing suggests that actually all have gaps of similar magnitude, and in such a case it is difficult to reduce the problem to a subset of the bands \cite{Sca14}. 

Symmetry aspects have always been useful, giving us robust arguments for the understanding of phases of matter and their stability. In particular for superconducting phases, Anderson's theorems \cite{And1,And2} give a qualitative understanding on the stability of different superconducting states in presence of key symmetry breaking fields. For example, singlet superconductivity is suppressed by the application of an external magnetic field, while triplet superconductivity is not robust in presence of certain inversion symmetry breaking fields. Recently \cite{Ram,Fis}, the authors showed that these criteria cannot be directly transferred to multi-orbital system, and introduce the concept of \emph{superconducting fitness} to address the question of stability of superconductivity in arbitrarily complex multi-orbital systems. With this tool in hand we can evaluate the stability of different gap structures in presence of inter-orbital effects and external symmetry breaking fields, leading to robust arguments to be contrasted with experiments.

The concept of superconducting fitness, $F(\bk)$ below, generalizes the idea of symmetry breaking encoded in Anderson's theorems. It does so by capturing the breaking of the degeneracy of the states to be paired, which can be understood as a band splitting. The concept is based on the modified commutator:
\begin{eqnarray}\label{SCFit}
F(\bk)(i\sigma_2)=H_0(\bk) \hat{\Delta}(\bk) - \hat{\Delta}(\bk)H^*_0(-\bk),
\end{eqnarray}
where $H_0(\bk)$ is the non-interacting Hamiltonian and $ \hat{\Delta}(\bk)$ is the normalized gap matrix, both in an orbital basis \cite{Ram}. $F(\bk)$ gives a direct measure of the incompatibility between an arbitrary pairing state and the normal state Hamiltonian. If $F(\bk)=0$ the gap matrix and the electronic structure are compatible, leading to a robust BCS-like instability towards a superconducting state, whereas if $F(\bk)$ is finite, there is some degree of incompatibility. This statement can be made explicit if we perturb the system by introducing $\delta H(\bk)$ to the non-interacting Hamiltonian and evaluate the superconducting fitness. In a single band picture, the instability of the superconducting phase under a perturbation can be directly seen as a reduction of the critical temperature following:
\begin{eqnarray}\label{Tc}
T_c&\sim& T_c^0\left(1 -\frac{7 \xi(3)}{64 \pi^2} \frac{ \left\langle  Tr|F(\bk )|^2  \right\rangle_{FS}}{(T_c^0)^2}\right).
\end{eqnarray}
Here $T_c$ and $T_{c0}$ are the critical temperatures in presence and absence of the perturbation, respectively, and $\langle...\rangle_{FS}$ denotes the average over the Fermi surface. Details on the derivation of this relation can be found in reference \cite{Ram}.

Applying this concept to \SRO we find that the most compatible triplet superconducting (SC) state has the d-vector oriented along the z-axis, as a consequence of the interplay of inter-orbital hopping and spin-orbit coupling \cite{Ram}. Concerning magnetic field effects, we were able to identify a new mechanism for the suppression of superconductivity in multi-orbital systems referred to as \emph{inter-orbital effect} (IEO) \cite{Ram}, in order to discriminate it from the standard orbital depairing mechanism. The IEO, together with the standard orbital depairing, is a good candidate to explain the unusual behaviour of the upper critical field in this material. 

In the next section we introduce the 3-band model for \SRO and review our previous results applying the concept of superconducting fitness to this system under magnetic field. We highlight some of the interesting experimental facts observed in strained \SRO \cite{Ste} in Sec.~\ref{SecStr} and extend our analysis for this new condition. We conclude and discuss further interesting directions for experiments in Sec.~\ref{SecCon}.

\section{Unstrained S\MakeLowercase{r}$_2$R\MakeLowercase{u}O$_4$}\label{SecUns}

\SRO is modelled by a multi-orbital Hamiltonian including the three low-lying $|{yz}\rangle$, $|{xz}\rangle$ and $|{xy\rangle}$ orbitals (here we label these as $|{1}\rangle$, $|{2}\rangle$ and $|{3}\rangle$, respectively). Based on the symmetries of the orbitals and the underlying lattice, one can construct a tight-binding model with hopping up to next-nearest neighbours and SOC. Introducing the six-dimensional Nambu basis:
\begin{eqnarray}\label{SROBasis}
\Psi\dg_{\bk} = (c\dg_{1\bk\up}, c\dg_{1\bk\down}, c\dg_{2\bk\up},c\dg_{2\bk\down}, c\dg_{3\bk\up}, c\dg_{3\bk\down}),
\end{eqnarray} 
the matrix form of the Hamiltonian is:
\begin{eqnarray}
H_0=\sum_{\bk} \Psi\dg_{\bk}
H_{SRO}(\bk)
\Psi_{\bk},
\end{eqnarray}
with
\begin{eqnarray}
H_{SRO}(\bk)=
\begin{pmatrix}
\epsilon_{1\bk} \sigma_0& t_\bk \sigma_0 +i \eta \sigma_3 & -i \eta \sigma_2\\
t_\bk \sigma_0 -i \eta \sigma_3 & \epsilon_{2\bk} \sigma_0 & i \eta \sigma_1\\
i \eta \sigma_2 & - i \eta \sigma_1& \epsilon_{3\bk} \sigma_0
\end{pmatrix},
\end{eqnarray}
where $\epsilon_{n\bk}$ is the dispersion for each of the orbitals originated from intra-orbital hopping, $t_\bk$ concerns inter-orbital hopping and $\eta$ is the magnitude of the SOC. Here $\sigma_0$ is the 2x2 identity matrix and $\sigma_{i}$ are Pauli matrices ($i=1,2,3$). The most important feature of the Hamiltonian above is its structure, which is determined by the character of the low-lying orbitals and the lattice symmetry. By symmetry, inter-orbital hopping is allowed only between orbitals $|{1}\rangle$ and $|{2}\rangle$, and the presence of SOC introduces mixing of all orbitals.

The gap matrix can also be written in the six-dimensional basis introduced in Eq.~\ref{SROBasis}. Considering only intra-orbital pairing it can be written as:
\begin{eqnarray}
\Delta(\bk) =
\begin{pmatrix}
\Delta_1(\bk) & 0& 0\\
0 &\Delta_2(\bk)  & 0\\
0 & 0 &\Delta_3(\bk) 
\end{pmatrix},
\end{eqnarray}
where
\begin{eqnarray}
\Delta_a(\bk) =
\phi_a \left(\mathbf{d}^a(\bk)\cdot \boldsymbol{\sigma} \right)(i\sigma_2) ,
\end{eqnarray}
for a triplet state, where the index $a=1,2,3$ refers to different orbitals, $\mathbf{d}^a(\bk)=(d^a_x(\bk), d^a_y(\bk), d^a_z(\bk))$ is a three-component vector and an odd function of momentum which parametrizes the order parameter and $ \boldsymbol{\sigma}=(\sigma_1,\sigma_2,\sigma_3)$; or
\begin{eqnarray}
\Delta_a(\bk) =
\phi_a d^a_0(\bk) (i\sigma_2) ,
\end{eqnarray}
for a singlet superconducting state, where the scalar function $d^a_0(\bk) $ is even in $\bk$. Here $\phi_a$ carries information about the magnitude and phase of the gap in orbital $a$ and $d^a_0(\bk)$ and $\mathbf{d}^a(\bk)$ are normalized basis functions which encode the momentum dependence of the gap.

\subsection{Magnetic field effects}\label{SecUnsMag}

The presence of external magnetic fields leads to time-reversal symmetry breaking and possible suppression of the SC state. 
Apart from the standard orbital depairing which is not included here, the effects of magnetic fields can appear in two different ways: by direct coupling to the spin degree of freedom as a Zeeman term or by coupling to the orbital angular momentum. For the specific case of \SRO\!\!, we can write the explicit matrix form for these effects in the multi-orbital basis introduced in Eq.~\ref{SROBasis}. We start with the Zeeman term:
\begin{eqnarray}
\delta H_{Z} 
&=& \begin{pmatrix}
 -\mathbf{h}\cdot \boldsymbol{\sigma} & 0 & 0\\
 0 & -\mathbf{h}\cdot \boldsymbol{\sigma} & 0\\
0 & 0 & -\mathbf{h}\cdot \boldsymbol{\sigma}
\end{pmatrix},
\end{eqnarray}
where $\mathbf{h}=(h_x,h_y,h_z)$ is the external magnetic field. Note that this perturbation is independent of the orbital character.  Focusing only on the field dependent part of the superconducting fitness, 
we find that the coupling of magnetic field to the spin degree of freedom leads to the same suppression of the critical temperature as in the single band case. Singlet SC is always suppressed in the presence of magnetic field, while triplet SC is suppressed if the magnetic field has a component in the same direction as the d-vector.

On the other hand, considering the coupling of the magnetic field to the orbital degree of freedom the perturbation has the form:
\begin{eqnarray}\label{Horb}
\delta H_{Orb}
&=& \begin{pmatrix}
0 & i h_{oz} \sigma_0 & - i h_{oy} \sigma_0\\
-i h_{oz} \sigma_0 & 0 & i h_{ox} \sigma_0 \\
i h_{oy} \sigma_0 &  -i h_{ox} \sigma_0 &0
\end{pmatrix}.
\end{eqnarray}
Here the orbital content is important to determine the structure of the matrix $\delta H_{Orb}$ and ultimately the effects of magnetic field on the superconducting state. Computing the superconducting fitness in order to verify the stability of different gap structures in the presence of orbital effects, we find, omitting the momentum dependence:
\begin{eqnarray}\label{FOS}
F_{Orb}&=& h_{ox} (\hat{\phi}^2 d_{j}^2+\hat{\phi}^3d_{j}^3) M_{7j}\\ \nonumber&-& h_{oy} (\hat{\phi}^1d_{j}^1+\hat{\phi}^3d_{j}^3) M_{5j} \\ \nonumber&+& h_{oz} (\hat{\phi}^1d_{j}^1+\hat{\phi}^2d_{j}^2) M_{2j},
\end{eqnarray}
where $M_{ij}=\lambda_i\otimes \sigma_j$, with $\lambda_i$ Gell-Mann matrices and $\sigma_i$ Pauli matrices. Here $j=0$ for singlet SC and $j=\{x,y,z\}$ for triplet SC. Note that for any gap structure among the different orbitals it is impossible to have all terms in $F_{Orb}$ equal to zero. The magnetic field is potentially detrimental to all SC states irrespective of its direction.  So we conclude that for multi-orbital systems the susceptibility of the critical temperature in presence of magnetic fields is not trivial and goes beyond the standard discussion of limiting effects based on symmetry arguments. We call this new effect \emph{inter-orbital effect} (IOE) in order to distinguish  it from the usual orbital depairing mechanism.

Concerning experiments, there are a two features that are still not completely understood in the upper critical field of \SRO\!\!:
\begin{itemize}

\item Anisotropy of the upper critical field: the angular dependence of the upper critical field does not follow the prediction of a pure anisotropic orbital effect described by an effective mass model (EMM) as expected for an ordinary layered type-II superconductor \cite{Deg,Yon13,Yon14};

\item  Transition becomes first order below $0.8K$ for fields within $2^o$ from the Ru-O plane, as recently observed by magnetocaloric effect \cite{Yon13} and specific heat measurements in ultra pure samples \cite{Yon14}. 

\end{itemize}

The presence of a first order transition might suggest that the Pauli paramagnetic limiting effect \cite{Ama15} is the dominant mechanism by which superconductivity is being suppressed. This would be inconsistent with NMR \cite{Ish98,Ish01,Mur} and neutron scattering \cite{Duf} experiments which show no observable change in the spin susceptibility through the superconducting transition for in-plane fields. An explanation for this unusual behaviour, compatible with all experimental results,  has not been established so far, and experimentalists have been suggesting that it is due to a new pair breaking mechanism \cite{Deg,Yon14}. In our recent work we suggest that the mechanism could be related to IOE, which can potentially account for \emph{both} the anisotropy and the change in the character of the transition, as we sumamrize in the following.

In type-II SC the orbital effect is the dominant pair breaking mechanism and superconductivity is suppressed in a second order phase transition \cite{Tin}.  In anisotropic materials this effect is captured by the EMM \cite{Yon13,Mor}:
\begin{eqnarray}\label{EMM}
h_{c2}^{EMM}(\theta) = \frac{h_{c2}(90^o)}{\sqrt{\sin^2\theta + \cos^2\theta/\Gamma^2 }},
\end{eqnarray}
where $\Gamma=\sqrt{M/m}$ is the anisotropy parameter, with $M$ the out of plane and $m$ the in-plane effective masses and $\theta$ is the angle the magnetic field relative to the conducting plane. As can be seen in Fig.~\ref{Angle} top, the EMM (blue, full line) describes the angular dependence of the upper critical field very well for angles larger than $2^o$, but for angles within $2^o$ of the plane there are clear deviations from the EMM \cite{Deg,Yon13,Yon14}. The extra suppression of the SC state in this region indicates that there should be a second mechanism which is most effective for in-plane fields. The standard paramagnetic limiting effect cannot account for this extra suppression of the SC state for in-plane fields if the d-vector is along the z-direction.

\begin{figure}[t]
\begin{center}
\begin{minipage}[c]{0.55\textwidth}
\includegraphics[width=\linewidth, keepaspectratio]{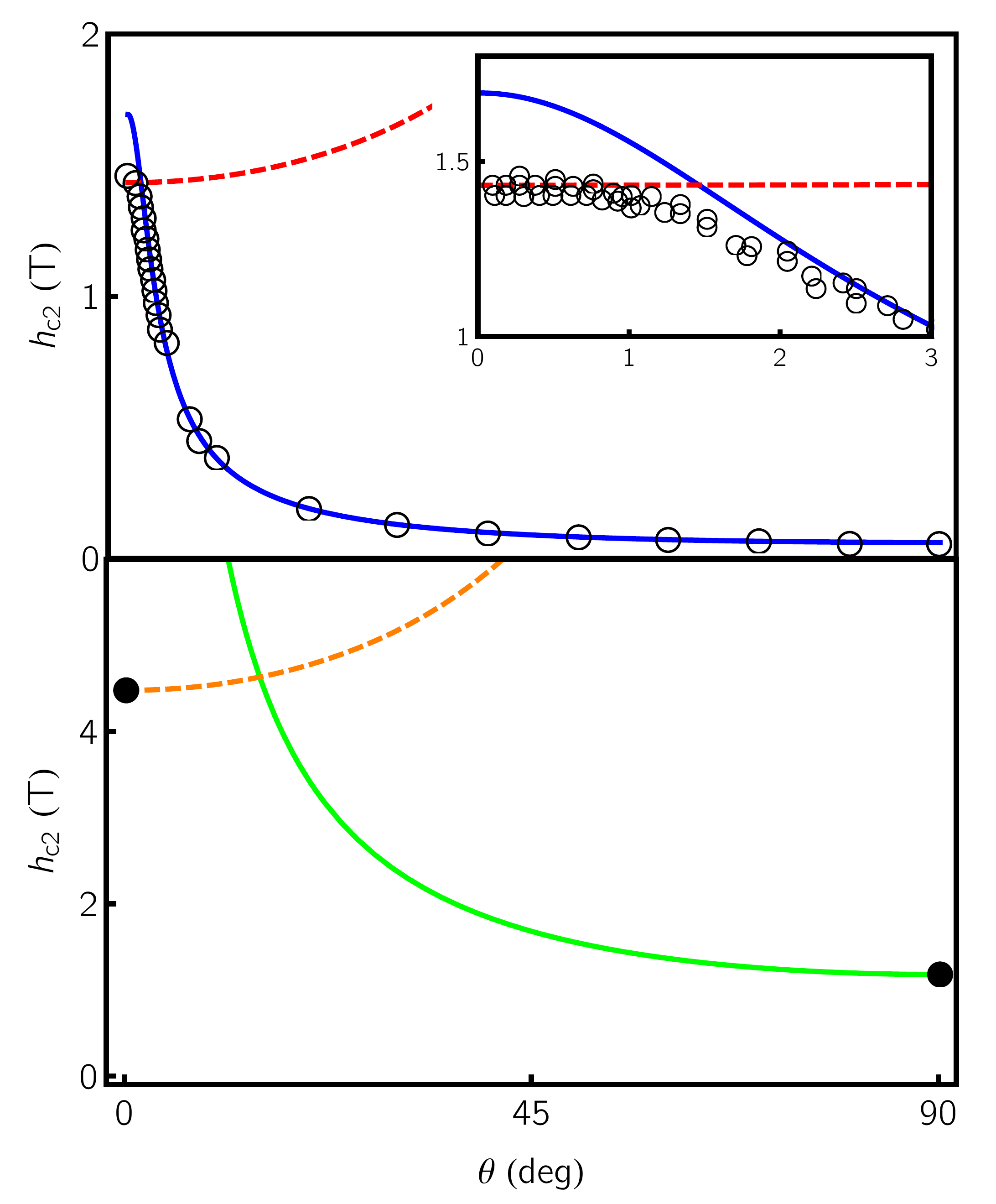}\label{Angle}
\end{minipage}
\begin{minipage}[c]{0.43\textwidth}
\caption{Angular dependence of the upper critical field. Top: Unstrained \SRO\!\!. The blue (full) line is the best fit for angles $\theta>2^o$ with the EMM of Eq.~\ref{EMM} with parameters $\Gamma=25$ and $h_{c2}(90^o)=0.07 T$.  The red (dashed) line refers to the IOE plotted phenomenologically as $h_{c2}^{IOE}= \frac{h_{c2}(0^o)}{\cos\theta}$, with $h_{c2}(0^o)=1.44 T$. The open circles are experimental data from Yonezawa et al. \cite{Yon13}. The inset shows the small angle region. Bottom: Strained \SRO\!. The green (full) line refers to the orbital effect described by the effective mass model EMM with parameters $\Gamma=25$ and $h_{c2}(90^o)=1.2T$.  The orange (dashed) line refers to the IOE with $h_{c2}(0^o)=4.5 T$. The full circles are experimental data from Steppke et al. \cite{Ste}.}
\end{minipage}
\label{Angle}
\end{center}
\end{figure}

Considering now the effects of the coupling of the magnetic field to the orbital angular momentum described by the IOE, we would expect a suppression of the critical temperature following Eq.~\ref{Tc} leading to a second order phase transition. In analogy to the standard paramagnetic limiting effect, the transition due to IOE is of first order and the upper critical field is in fact determined by comparing the changes in the magnetic and condensation energies across the superconducting transition. The condensation energy can be estimated as:
\begin{eqnarray}
E_C=-\frac{N(0) |\Delta|^2}{2},
\end{eqnarray}
where $N(0)$ is the density of states at the Fermi level and $|\Delta|$ the magnitude of the superconducting gap. In the IOE, the magnetic energy comes from the polarizarion of the orbitals within a given band, so here we use $\Delta\chi_{Orb}$ as the change in the orbital susceptibility from the normal to the SC state:
\begin{eqnarray}
E_M=-\frac{\Delta\chi_{Orb}h_x^2}{2},
\end{eqnarray}
which leads to the critical field
\begin{eqnarray}
h_{c2}^{IOE}(\theta) \approx \sqrt{\frac{N(0)}{\Delta\chi_{Orb}}} \frac{T_{c}^0}{\cos\theta}.
\end{eqnarray}

Here we would like to emphasize that the unusual angular dependence of the upper critical field as shown in Fig.~\ref{Angle} can be understood if both mechanisms, standard orbital and inter-orbital, are present. Moreover, we have seen that the standard EMM would lead to a second order transition, while depairing following from the field-induced inter-orbital coupling favours a discontinuous transition much analogous to the expectations for paramagnetic limiting. Therefore we believe that inter-orbital coupling for in-plane fields provides a strong candidate to explain the unusual suppression of superconductivity seen in experiments on \SRO\!\!.  

\section{Magnetic field effects in S\MakeLowercase{r}$_2$R\MakeLowercase{u}O$_4$ under strain}\label{SecStr}

We now turn to the analysis of the upper critical field for strained \SRO\!\!. Under uniaxial stress \SRO can have its critical temperature enhanced up to $3.4K$ \cite{Ste} (from the original $1.5K$ for the unstrained material \cite{Mac}). Compared to the unstrained sample, there are a few quantitative changes on the upper critical field of the strained sample which we highlight below:

\begin{itemize}

\item Larger critical fields: compared to the unstrained material, the critical field along the z-axis is enhanced by a factor of $\sim 20$, while the in-plane critical field is enhanced by a factor of $\sim 3$ \cite{Ste,Deg,Yon13,Yon14}. Given the fact that the critical temperature was enhanced by a factor of $\sim 2.3$, an enhancement of the upper critical field is expected, but the rather different factors for different orientations is unusual;

\item Less anisotropy: the ratio of the upper critical field (in-plane/z-axis) for unstrained \SRO is around $25$ \cite{Deg,Yon13,Yon14}, while for the strained sample it is approximately $4$ \cite{Ste}. It is difficult to understand this dramatic reduction of the anisotropy purely within orbital effects by a reduction in the mass anisotropy. The strong 2D character for the strained sample is observed by experiments based on the very different initial slope of $h_{c2}$ versus temperature near $ T_c $ for fields along the plane and the z-axis \cite{Ste}. It is also corroborated by recent DFT calculations for this system under strain \cite{Ste};

\item The transition turns to first order at higher temperature: for unstrained SRO the transition becomes first order for in plane fields only below $0.8K$ \cite{Yon13,Yon14}. For the strained case the transition is first order below $1.8K$ \cite{Ste}. This indicates that the mechanism leading to the first order transition takes over at larger energy scales.

\end{itemize}

The presence of both the standard orbital effect and the IOE can still be in agreement with the findings of the upper critical field under strain. So far we do not have access to the entire angular dependence of the upper critical field, but we can focus on the two extreme points from Steppke et al. \cite{Ste}. Assuming that the dominant mechanism for suppression of superconductivity with fields along the z-axis is the standard orbital effect which is captured by the EMM, and that under strain the effective mass ratio is not strongly modified (so we still assume $\Gamma\sim 25$ as for the unstrained sample), we can plot the angular dependence in Fig.~\ref{Angle} bottom, green full line. Note that the prediction for the in-plane upper critical field overestimates the experimental value by factor of approximately $7$, much larger than in the unstrained case discussed above. 

Assuming orbital effects are dominant for fields along the z-axis, in order to have agreement with the value of the upper critical field along the z-axis we choose the value $h_{c2}(90^o)=1.2T$ at $ T= 0 K$ for the plot, much larger the the value for the unstrained sample. Within Ginzburg-Landau theory $h_{c2}= \phi_0/2\pi \xi^2$, so an enhancement of the z-axis upper critical field by a factor of $20$ indicates a reduction in the coherence length by a factor of approximately $4.5$. From BCS theory the coherence length can be written as $\xi\approx \hbar v_F/\pi T_c$, so the reduction of the coherence length is consistent with the enhancement of the critical temperature. The critical temperature increases only by a factor of $2.3$. But given that this estimate is rather crude (it does not take into account changes in the Fermi velocity, for example), we believe this argument is consistent with the fact that we still have the suppression of superconductivity by the standard orbital effect for fields along the z-axis.

The presence of a first order transition for in-plane fields in the strained sample again suggests the presence of Pauli paramagnetic limiting effects, and naively of a singlet superconducting state. In the previous section we review our recent work which points out that the IEO can also account for such a phenomenology with a triplet superconducting state \cite{Ram}. We believe this picture to be more appealing since it is in agreement with other experiments that indicate the presence of a triplet state for the unstrained sample. An extension of this idea to the strained case would be natural if there is no phase transition under strain.

Concerning the in-plane upper critical field, from BCS theory, the standard Pauli limiting effect predicts $h_{c2}/T_c\approx 1.25$ \cite{Tin,Sig08}. For magnetic field along the z-axis experiments show that $h_{c2}/T_c\approx 1 $ for the unstrained sample and $h_{c2}/T_c\approx 1.3 $ for the strained sample, indicating that a Pauli limiting-like effect (with $h_{c2}$ scaling linearly with $T_c$) is consistent with the observations. We plot in Fig.~\ref{Angle} bottom, orange dotted line, the respective IEO. We find that this mechanism overcomes the standard orbital effect for much larger angles (at $\theta \approx 20^o$). This reflects that fact that a smaller magnitude of in-plane magnetic field is necessary for suppression of superconductivity. From another perspective it indicates that this mechanism taking over at higher temperatures, being in accordance with the observed first order transition over an extended temperature range for the strained sample. Also, the fact that this mechanism dominates the behaviour of $h_{c2}$ for small angles dramatically reduces the anisotropy in comparison to the expected value due to the mass anisotropy. In conclusion, the IEO is again able to consistently put together two puzzling features of the upper critical field.

\section{Conclusion}\label{SecCon}

Here we would like to emphasize that one can understand the phenomenology of the upper critical field in \SRO assuming a triplet superconducting state. This is possible based the recently discussed mechanism for the suppression of superconductivity in multi-orbital systems called inter-orbital effect. The presence of a first order phase transition under a magnetic field suggests that the Pauli paramagnetic limiting effect is at play, and a naive analysis would lead us to think that a singlet superconducting state is the only possibility which is compatible with the observations. We have argued that the IEO can also lead to first order phase transitions under external magnetic fields by the polarization of the orbital content within each band, in analogy to the spin polarization in the Pauli paramagnetic limiting mechanism. Furthermore, a quantitative analysis of the changes in the values of the upper critical fields and critical temperatures leads to a consistent picture for the unstrained and uniaxially strained samples. In particular, the two extreme angles are dominated by different mechanisms such that the upper critical field scales differently with the critical temperature.

Our discussion does not rule out the possibility of a phase transition from triplet to singlet superconductor as a function of strain. In that case a discontinuity or an inflection point in $h_{c2}$ as a function of strain should be observed experimentally. We believe that experiments along this line would be of great interest for a better understanding of the superconducting phase of \SRO\!\!.  It would also be important to follow the complete angular dependence of the anisotropy in order to observe at which angles the deviations of the EMM appear in the strained samples.  If such an experiment is realized one could also determine by how much the anisotropy parameter $\Gamma$ changes in comparison to the unstrained sample, so we can understand if the reduction of anisotropy is due to the change in the effective masses or due to the IOE taking over at larger angles, as we discuss above.

\section*{Acknowledgements}
We thank M. H. Fischer, K. Ishida and C. W. Hicks for stimulating discussions. This work was supported by Dr. Max R\"{o}ssler, the Walter Haefner Foundation and the ETH Zurich Foundation (AR) and by the Swiss National Science Foundation (MS).


\end{document}